\documentclass[twocolumn,showpacs,superscriptaddress,showkeys,reprint,amsmath,amssymb,aps,prb]{revtex4-1}
\usepackage{graphicx,color}
\usepackage{dcolumn}
\usepackage{hyperref}
\usepackage{bm}
\usepackage{epstopdf}
\usepackage{amsmath}
\hfuzz=\maxdimen
\usepackage{float}

\usepackage{appendix}

\begin{document}

\title{Evidence the ferromagnetic order on CoSb layer of LaCoSb$_2$}

\author{Muyuan Zou}
\affiliation{State Key Laboratory of Surface Physics and Department of Physics, Fudan University, Shanghai 200433, China}

\author{Jianan Chu}
\affiliation{State Key Laboratory of Functional Materials for Informatics, Shanghai Institute of Microsystem and Information Technology, Chinese Academy of Sciences, Shanghai 200050, China}
\affiliation{University of Chinese Academy of Sciences, Beijing 100049, China}

\author{Hui Zhang}
\affiliation{State Key Laboratory of Functional Materials for Informatics, Shanghai Institute of Microsystem and Information Technology, Chinese Academy of Sciences, Shanghai 200050, China}

\author{Tianzhong Yuan}
\affiliation{State Key Laboratory of Surface Physics and Department of Physics, Fudan University, Shanghai 200433, China}

\author{Peng Cheng}
\affiliation{Department of Physics, Renmin University of China, Beijing 100872, China}

\author{Wentao Jin}
\affiliation{Key Laboratory of Micro-Nano Measurement-Manipulation and Physics (Ministry of Education), School of Physics, Beihang University, Beijing 100191, China}

\author{Da Jiang}
\affiliation{State Key Laboratory of Functional Materials for Informatics, Shanghai Institute of Microsystem and Information Technology, Chinese Academy of Sciences, Shanghai 200050, China}
\affiliation{Center of Materials Science and Optoelectronics Engineering, University of Chinese Academy of Sciences, Beijing 100049, China}

\author{Xuguang Xu}
\affiliation{School of Physical Science and Technology, ShanghaiTech University, Shanghai 201210, China}

\author{Wenjie Yu}
\affiliation{State Key Laboratory of Functional Materials for Informatics, Shanghai Institute of Microsystem and Information Technology, Chinese Academy of Sciences, Shanghai 200050, China}

\author{Zhenghua An}
\affiliation{State Key Laboratory of Surface Physics and Department of Physics, Fudan University, Shanghai 200433, China}
\affiliation{Collaborative Innovation Center of Advanced Microstructures, Nanjing University, Jiangsu 210093, China}

\author{Xinyuan Wei}
\email{xinyuanwei@fudan.edu.cn}
\affiliation{State Key Laboratory of Surface Physics and Department of Physics, Fudan University, Shanghai 200433, China}

\author{Gang Mu}
\email{mugang@mail.sim.ac.cn}
\affiliation{State Key Laboratory of Functional Materials for Informatics, Shanghai Institute of Microsystem and Information Technology, Chinese Academy of Sciences, Shanghai 200050, China}

\author{Wei Li}
\email{w$_$li@fudan.edu.cn}
\affiliation{State Key Laboratory of Surface Physics and Department of Physics, Fudan University, Shanghai 200433, China}
\affiliation{Collaborative Innovation Center of Advanced Microstructures, Nanjing University, Jiangsu 210093, China}

\date{\today}

\begin{abstract}
The emergence of unconventional superconductivity is generally considered to be related to spin fluctuations. Unveiling the intriguing behaviors of spin fluctuations in parent compounds with layered transition-metal ions may shed light on the search for exotic unconventional superconductors. Here, based on the framework of the first-principles calculations, we theoretically propose that LaCoSb$_2$ is a weak antiferromagnetic layered metal with an in-plane ferromagnetic moment of 0.88 $\mu_B$ at the Co sites, as a candidate parent compound of the cobalt-based superconductors. Importantly, this theoretical finding is experimentally supported by our magnetization measurements on polycrystalline samples of LaCo$_{0.78}$Sb$_2$. Following the symmetry analysis, we suggest a possible $p$-wave superconductivity hosted in doped LaCoSb$_2$ emerging at the verge of ferromagnetic spin fluctuations, which implies potential applications in topological quantum computing in future.
\end{abstract}

\maketitle

\section{Introduction}

Studies on topological superconductors~\cite{Qi2011,Sato2017,Beenakker2013,Flensberg} have triggered enormous research interest during recent years, owing to the fact that these systems host Majorana quasiparticles~\cite{Alicea,Franz2015}, which obey non-Abelian statistics and can be used to encode and manipulate quantum information in a topologically protected manner~\cite{Nayak2008,Kitaev}. In solid-state systems, $p$-wave superconductors~\cite{Ivanov2001,Aguado} or $\nu=\frac{5}{2}$ fractional quantum Hall states~\cite{Green2000,Moore1991,Read1992} are considered to exhibit Majorana quasiparticles. The first candidate chiral $p$-wave superconductor is Sr$_2$RuO$_4$ evidenced by previous NMR measurements~\cite{Maeno1994,Maeno1998}, which breaks the time-reversal symmetry~\cite{JXia2006,Kapitulnik2009}. However, a recent NMR measurement contradicts the previous finding due to the heat-up effect in the Knight shift~\cite{Pustogow2019,Ishida2019}.

Alternatively, a theoretical study predicted the appearance of Majorana bound states at the interface between a topological insulator and a conventional superconductor~\cite{LFu2008}. Later, a generic theory which does not involve any topological insulator is also proposed, claiming that the combination of Rashba spin-orbit coupling, the Zeeman effect, and the proximity effect between an $s$-wave superconductor and a two-dimensional metal will induce effective $p$-wave superconductivity~\cite{Sau2010}. Experimentally, evidence of Majorana bound states has been reported in various systems, including one-dimensional nanowires in contact with superconductors~\cite{Mourik2012,MDeng2012,ADas2012,Deng2016,HZhang2018}, at the edges of iron atom chains formed on the surface of superconducting lead~\cite{Perge2014}, at the interface between a topological insulator and an $s$-wave superconductor~\cite{JFJia2015,JFJia2016}, quantum spin liquids~\cite{Banerjee2016}, 
and iron-based superconductors~\cite{DLFeng,JXYin2015,HHWen,Wang2018,SZhu2019,CChen2020}.

\begin{figure*}
\centering{}\includegraphics[bb=10 10 600 300,width=15cm,height=7cm]{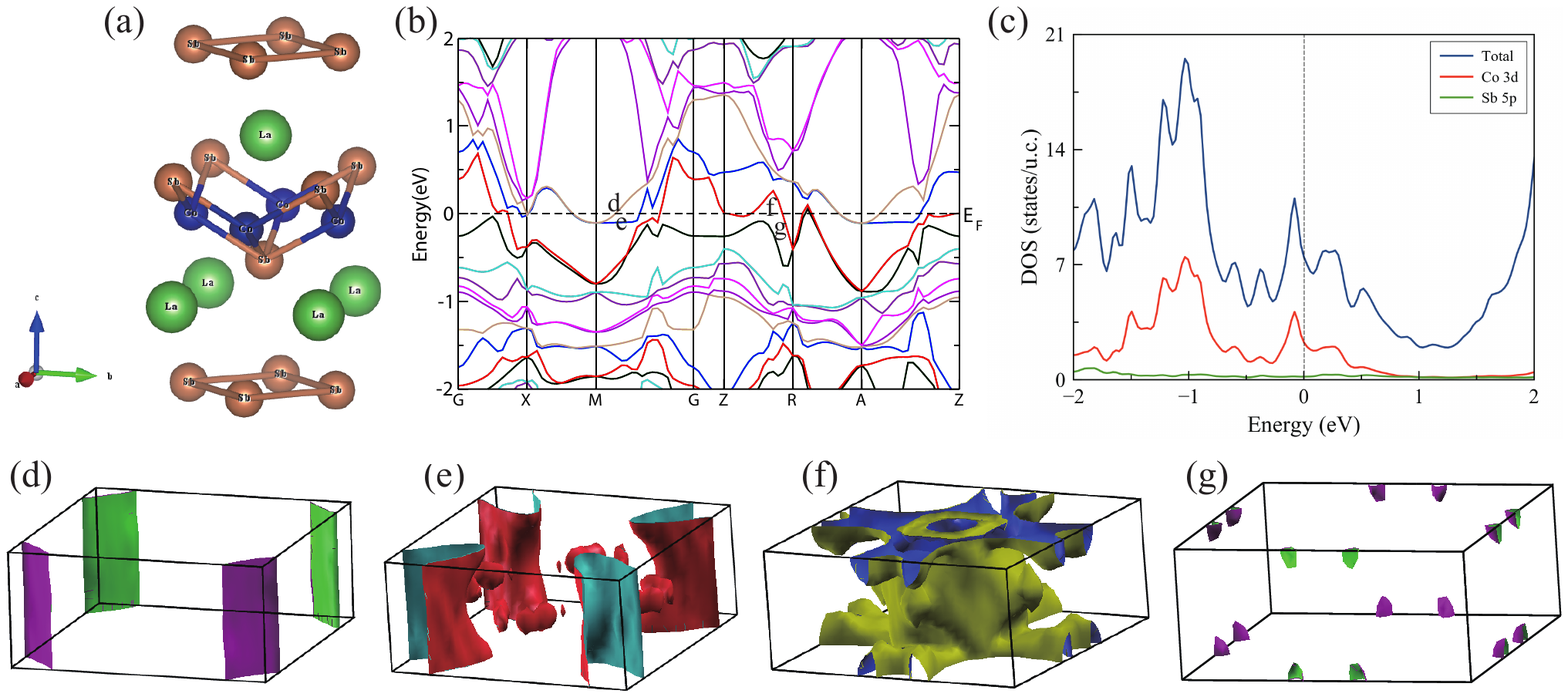}
\caption{(Color online) (a) Schematic illustration of the crystal structure of LaCoSb$_2$ containing a sandwiched CoSb layer. (b) Electronic band structure, and (c) total DOS and PDOS on Co $3d$ and Sb $5p$ orbitals of the NM state of LaCoSb$_2$. (d)-(g) Fermi surface topologies for the NM state of LaCoSb$_2$ with the corresponding labels (``d, e, f, g'') shown in (b). Fermi energies are set to zero.}
\label{fig1}
\end{figure*}

In this paper, we theoretically propose doped LaCoSb$_2$ as a possible candidate topological superconductor based on the symmetry analysis that the superconductivity might emerge at the verge of ferromagnetic (FM) spin fluctuations on CoSb layers. LaCoSb$_2$~\cite{Rogl1994}, as shown in Fig.~\ref{fig1}(a), has a similar crystal structure to 112-type iron-based superconductors~\cite{Katayama2013,Yakita2014}. Analogous to the electronic properties of layered iron-based superconductors~\cite{WLi,WLi2012}, the conducting electrons in LaCoSb$_2$ mainly come from the contribution of Co $3d$ states partially hybridized with the Sb $5p$ states on CoSb layers. Previous first-principles calculations demonstrate that the nonmagnetic (NM) state of LaCoSb$_2$ displays the behaviors of a high-symmetry line semimetal~\cite{TZhang2019}. Surprisingly, a recent theoretical prediction reveals a possible realization of high-transition temperature superconductivity in a freestanding or SrTiO$_3$-supported CoSb monolayer~\cite{Zhang2020}. Experimentally, monolayer CoSb has been successfully grown on SrTiO$_3$(001) substrates by molecular beam epitaxy, in which symmetric superconducting gaps with coherence peaks at around $\pm$ 6 meV close to the Fermi level were observed by $in$ $situ$ scanning tunneling spectroscopy~\cite{CDing2019}, that is, three times smaller compared with monolayer FeSe on SrTiO$_3$(001) substrates~\cite{QYWang2012}. Furthermore, $ex$-$situ$ magnetization measurements demonstrate the superconducting transition temperature of $T_c\approx$ 14 K for CoSb/SrTiO$_3$, accompanied by a weak net FM moment~\cite{CDing2019}. Interestingly, for LaCoSb$_2$, our first-principles calculations find a weak antiferromagnetic (AFM) ground state with an in-plane FM moment of 0.88 $\mu_B$ at Co sites. Additionally, our magnetization measurements support the presence of AFM spin fluctuations in polycrystalline samples of LaCo$_{0.78}$Sb$_2$ with the magnetic moment of 0.78 $\mu_B$, in good agreement with the theoretical findings. Thus, LaCoSb$_2$ becomes a candidate parent compound of cobalt-based superconductors, as the FM order can be suppressed gradually by chemical doping and the superconductivity might emerge at the verge of FM spin fluctuations, making doped LaCoSb$_2$ a possible $p$-wave superconductor based on the group analysis. 

The rest of this paper is organized as follows. In Sec.~\ref{Sec2}, we present the electronic and magnetic structures of LaCoSb$_2$ using first-principles calculations. The experimental magnetization measurements on polycrystalline LaCo$_{0.78}$Sb$_2$ are shown in Sec.~\ref{Sec3}. The symmetry analysis on superconducting gap structures is addressed in Sec.~\ref{Sec4}. Finally, we give a brief conclusion in Sec.~\ref{Sec5}.

\section{The First-Principles Calculations}\label{Sec2}

The calculations in this work are performed using the all-electron full potential linear augmented plane wave plus local orbitals (FP-LAPW+lo) method~\cite{Singh} as implemented in the WIEN2k code~\cite{wien2k}. The exchange-correlation potential is calculated using the generalized gradient approximation as proposed by Perdew, Burke, and Ernzerhof~\cite{PBE1996}. Since the spin-orbit coupling strength is proportional to $Z^4$ (where $Z$ is the atomic number; $Z=51$ for Sb)~\cite{Li2014}, the spin-orbit coupling is nonnegligible for LaCoSb$_2$. Therefore, the spin-orbit coupling is included with the second variational method throughout the calculations. Furthermore, a 1000 $\mathbf{k}$-point is chosen to ensure calculations with an accuracy of $10^{-5}$ eV, and all crystallographic parameters determined experimentally~\cite{Rogl1994} including the lattice constants and the internal coordinates are adopted in the calculations [Fig.~\ref{fig1}(a)]. The stability of the crystal structure of LaCoSb$_2$ is shown in Fig.~\ref{A1} in Appendix~\ref{SecA1}. 

First, we focus on the NM state behavior of LaCoSb$_2$, which means that no spin polarization is allowed on the Co ions in calculations. Such a study can provide a reference for examining whether the magnetic ordered state is favorable. Fig.~\ref{fig1}(b) shows the electronic band structure of the NM state of LaCoSb$_2$, which has much more complicated electronic dispersive features than that for the 112-type iron-based superconductor~\cite{Katayama2013}. In contrast to the five bands across the Fermi level in iron-based superconductors, however, there are mainly three of the four bands crossing the Fermi level contributing to electron conduction in LaCoSb$_2$, which can be seen clearly in the Fermi surface topologies shown in Figs.~\ref{fig1}(d)-(g). Verifying the orbital characteristics of the energy bands around the Fermi surface (see Fig.~\ref{A2} in Appendix~\ref{SecA2}), we note the prominent contributions stemming from the Co $t_{2g}$ ($d_{x^2-y^2}$, $d_{xz}$, and $d_{yz}$) orbitals partially associated with the $d_{z^2}$ orbital. From the viewpoint of crystal field theory~\cite{WLi2012}, the Co ions are coordinated by the Sb tetrahedron, and the crystal field will normally split the five Co $3d$ orbitals into low-lying twofold $e_g$ ($d_{xy}$ and $d_{z^2}$) orbitals and up-lying threefold $t_{2g}$ orbitals opposite the octahedral case. Taking the Coulomb interaction into account, the actual instance by analyzing the orbital dependent energy band structure is opposite to that we expect from a simple tetrahedron crystal field: the low-lying manifold is threefold ($t_{2g}$) and the up-lying manifold is twofold ($e_g$). For the Co$^{3+}$ ion, the nominal number of $3d$ electrons is six. The electron conduction around the Fermi level mainly arises from the low-lying $t_{2g}$ orbitals, and partly arises from the $e_g$ orbitals due to the strong Hund's coupling effect, similar to that in iron-based superconductors~\cite{Singh2008,WLi2012_EPL}. Since the atomic radius of Sb is much larger, it drives the enhancement of the interlayer covalent bonding, resulting in the strong three-dimensionality of Fermi surface topologies, as shown in Figs.~\ref{fig1}(e)-(g). Additionally, it is interesting to point out that there is an electron-like pocket with a nearly cylindrical shape located around the $M$ point, as shown in Fig.~\ref{fig1}(d), similar to the electron pocket existing in the 112-type iron-based superconductor~\cite{Katayama2013}, suggesting a strong two-dimensional electronic behavior. Furthermore, previous first-principles calculations have demonstrated that the band structure of the NM state of LaCoSb$_2$ displays the behaviors of a high-symmetry line semimetal~\cite{TZhang2019}, consistent with our present calculations.

\begin{table}
\caption{The calculated total energy of various magnetically ordered states of LaCoSb$_{2}$. $\Delta E$ is the total energy difference per Co atom with respect to the NM state, and $m_{Co}$ is the local magnetic moment per Co atom.}
\begin{ruledtabular} %
\begin{tabular}{ccccc}
LaCoSb$_{2}$  & FM & AFM & A-AFM & c-AFM   \\
\hline
$\Delta E$ (meV/Co)&-80.86 &-3.54 &$\mathbf{-81.68}$& -37.15 \\
\hline
$m_{Co}$($\mu_{B}$)& 0.87  & 0.30& $\mathbf{0.88}$  & 0.69 \\
\end{tabular}\end{ruledtabular}
\label{tabel1}
\end{table}

The calculated density of states (DOS) and the projected DOS (PDOS) on Co $3d$ and Sb $5p$ orbitals of LaCoSb$_2$ are shown in Fig.~\ref{fig1}(c). It can be seen that the conduction electrons mainly come from the contribution of Co $3d$ states partially hybridized with Sb $5p$ states. Verifying the value of the DOS at the Fermi level, $N(E_f)=2.3$ states per eV per Co atom, we notice that this value is much larger than that in iron-based superconductors~\cite{WLi2012}. While magnetism may occur with lower values of the DOS, it must occur within a band picture if the Stoner criterion~\cite{Singh2008}, $N(E_f)\times I > 1$, is met, where $I$ is the Stoner parameter, taking values of $0.7-0.9$ eV for ions near the middle of the $3d$ series (note that the effective $I$ can be reduced by hybridization)~\cite{Singh2008}, suggesting that the NM state is unstable against the magnetic states for LaCoSb$_2$.

To explore the ground state with magnetic ordering of LaCoSb$_2$, we have calculated four possible magnetically ordered states in the Co layer with FM, N\'{e}el-AFM, and A-type AFM (A-AFM)~\cite{Wollan} orders, as well as collinear AFM (c-AFM) order (align FM order along one direction and AFM order along the other direction in the Co-Co square-lattice layer, similar to that in LaOFeAs~\cite{Dong2008,Dai2008}). The corresponding total energies of various magnetic ordering states are listed in Table~\ref{tabel1}. It is shown that A-AFM order with FM order in Co layers and AFM order between Co layers is the lowest energetic state for LaCoSb$_2$. The calculated magnetic moment is 0.88 $\mu_B$ on Co atoms.

\begin{figure}[tbp]
\centering{}\includegraphics[bb=15 25 560 240,width=8.2cm,height=3.2cm]{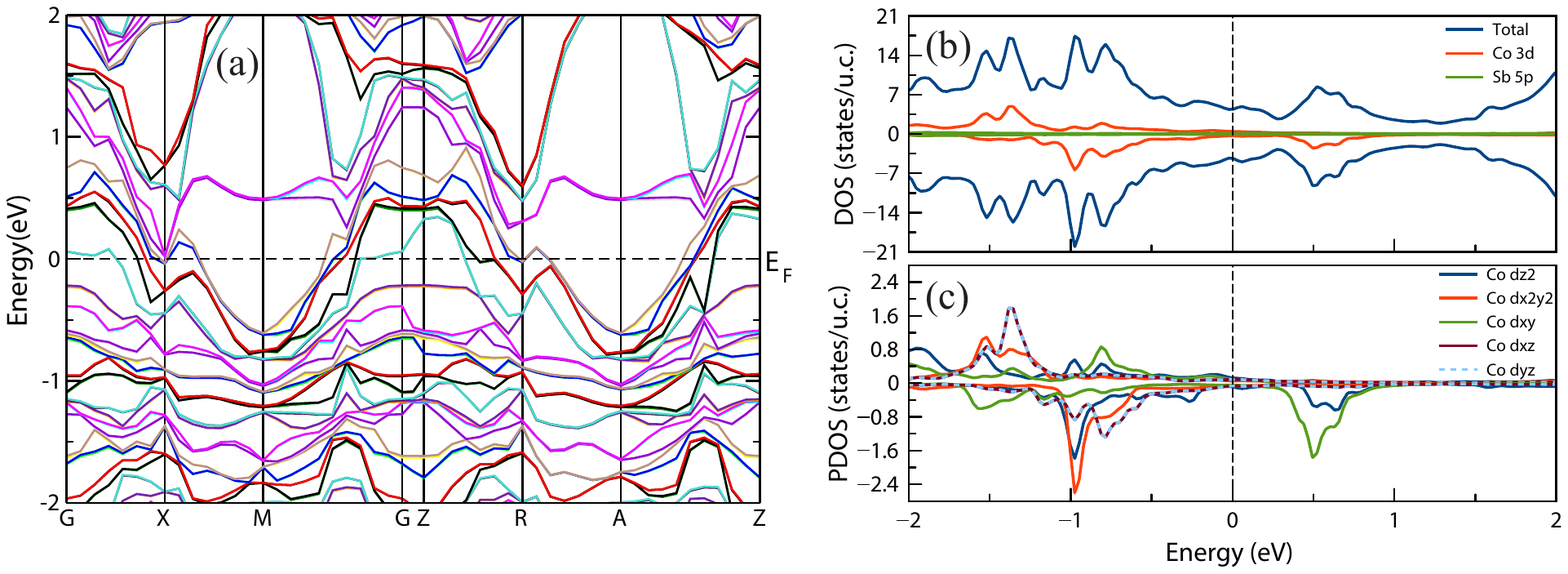}
\caption{(Color online) (a) Band structure, (b) total DOS, and (c) PDOS on one of the Co $3d$ and Sb $5p$ orbitals for the A-AFM ordered state of LaCoSb$_2$. Fermi energies are set to zero.}
\label{fig2}
\end{figure}

The calculated low-energy band structure, the corresponding total DOS and the PDOS on Co $3d$ and Sb $5p$ orbitals, and the orbital resolved PDOS on one of the Co $3d$ states, with the A-AFM ordered state in LaCoSb$_2$, are shown in Fig.~\ref{fig2}. Compared with the NM state shown in Fig.~\ref{fig1}, we find that most of the states around the Fermi level are gapped by the A-AFM order. The corresponding electronic DOS at the Fermi level is $N(E_f)=0.45$ states per eV per Co atom, which is significantly less than that of the NM state (2.3 states per eV per Co atom), as intuitively expected. Furthermore, the calculated orbital dependent PDOS in Fig.~\ref{fig2}(c) suggests that the Co $e_g$ orbitals are mainly responsible for the electron conduction (also see details in Fig.~\ref{A3} in Appendix~\ref{SecA3}). This result is also consistent with our discussion above within the framework of crystal field theory. Clarifying the role of the magnetic spin fluctuations may further provide insight into the mechanism of superconductivity that is possibly driven by electron-electron correlation.

In order to quantify the magnetic interactions for revealing microscopic spin fluctuation effects on LaCoSb$_2$, we consider a phenomenological theoretical Heisenberg model on the Co atoms as~\cite{ZZhou2019}:
\begin{equation}
\hat{H} = J_1\sum_{\langle i,j\rangle}\vec{S}_i\vec{S}_j+J_{2}\sum_{\langle\langle i,j\rangle\rangle}\vec{S}_i\vec{S}_j+J_{\perp}\sum_{\langle i,j\rangle}\vec{S}_i\vec{S}_j,
\label{Eq1}
\end{equation}
where $\vec{S}$ is the magnitude of Co spin, and $\langle i,j\rangle$ and $\langle\langle i,j\rangle\rangle$ denote the summation over the nearest-neighbor and next-nearest-neighbor sites, respectively. The parameters $J_1$ and $J_2$ describe the nearest-neighboring and next-nearest-neighboring intralayer exchange interactions, respectively, and $J_{\perp}$ denotes the nearest-neighboring interlayer exchange interaction. From the calculated energies for various magnetic configurations listed in Table~\ref{tabel1}, the magnetic exchange couplings $J_1=-19.33$ meV, $J_2 =-1.26$ meV, and $J_{\perp}= 0.41$ meV are found for LaCoSb$_2$, suggesting FM order in the Co layer and weak AFM interlayer coupling. The strong intralayer FM order mainly originates from the superexchange interaction with strong Hund's coupling mediated by Sb $5p$ orbitals~\cite{Maekawa2004}. Importantly, it is noteworthy that FM order on CoSb layers has also been observed by magnetization measurements in monolayer films of CoSb grown on SrTiO$_3$ substrates by molecular beam epitaxy, which supports our theoretical findings, although the nature of FM was explained there to result from the possible tellurium capping layer~\cite{CDing2019}. Furthermore, we also provide our magnetization data in the following section to further support the theoretical prediction of the presence of magnetic ordering in the bulk system of LaCoSb$_2$.

\begin{figure}
\includegraphics[width=8.5cm]{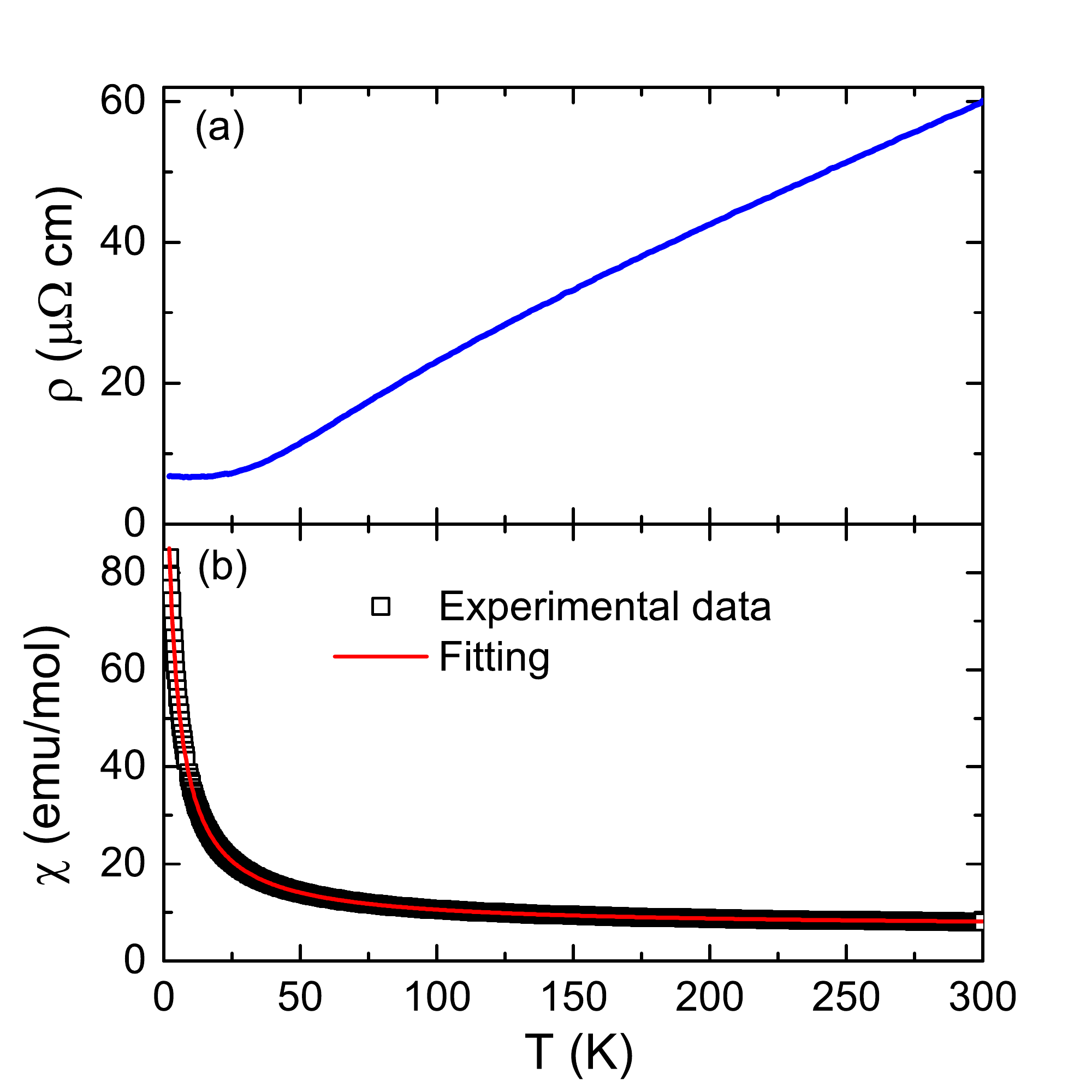}
\caption{(Color online) (a) Temperature dependence of the electric resistivity of LaCo$_{0.78}$Sb$_2$. (b) Temperature dependence of the dc magnetic susceptibility of LaCo$_{0.78}$Sb$_2$ with the applied magnetic field of 0.5 T.}
\label{fig3}
\end{figure}

\section{Experimental measurements}\label{Sec3}

We experimentally synthesized LaCo$_{0.78}$Sb$_2$ polycrystalline samples by an arc-melting method as described in the literature~\cite{Rogl1994}, whose crystal structure is shown in Fig.~\ref{fig1}(a) and XRD pattern is shown in Fig.~\ref{A4} in Appendix~\ref{SecA4}, to verify the theoretical findings of the magnetic behaviors in the bulk system of LaCoSb$_2$. In Fig.~\ref{fig3}(a), we show the electronic transport data on LaCo$_{0.78}$Sb$_2$ as a function of the temperature measured using the Physical Property Measurement System (PPMS DynaCool Quantum Design). The electric resistivity decreases monotonously with decreasing temperature, indicating typical metallic behavior for LaCo$_{0.78}$Sb$_2$. This is in good agreement with our first-principles calculations shown in Fig.~\ref{fig2}. Additionally, it is noteworthy that the magnitude of electric resistivity is about 1-2 orders lower than that of the iron-based superconductors~\cite{Mu2009}. Furthermore, we have also measured the dc magnetic susceptibility of LaCo$_{0.78}$Sb$_2$ as a function of the temperature using a magnetic property measurement system (Quantum Design MPMS3), which displays a monotonous increase with decreasing temperature, as shown in Fig.~\ref{fig3}(b), suggesting the absence of long-range magnetic ordering down to 2 K. This behavior is fitted by a generalized Curie-Weiss (CW) law~\cite{Amornpitoksuk2008},
\begin{equation}
\chi(T) = \chi_0+\frac{C}{T-\theta_{CW}},
\label{Eq2}
\end{equation}
where $\chi_0$ is the temperature-independent contribution that accounts for core diamagnetism and Van Vleck paramagnetism, while the second term is the CW law with Curie constant $C=\frac{N_A\mu_{eff}^2}{3k_B}$ and CW temperature $\theta_{CW}$. The fitting curve is shown in Fig.~\ref{fig3}(b) by the solid red line, indicating good validity of the CW fitting. Through CW fitting, the obtained $\theta_{CW}$ value is $-2.9$ K. The negative value of $\theta_{CW}$ suggests the existence of AFM spin fluctuation in LaCo$_{0.78}$Sb$_2$. On the mean-field level, the CW temperature can be expressed as $\theta_{CW}=-z\tilde{J}_{\perp}|S|^2/3k_B$, where $z$ is the number of nearest-neighbor spins along the $c$ axis and $k_B$ is the Boltzmann constant~\cite{bookQTM}. For the layered lattice system, the number of nearest neighbors along the $c$ axis is $z=2$ for LaCoSb$_2$, which yields an AFM exchange interaction of $\tilde{J}_{\perp}\simeq0.38$ meV, in good agreement with the theoretically calculated $J_{\perp}=0.41$ meV. From the fitting value of $C = 382$ emu$\cdot$K/mol, we derive the effective magnetic moments $\mu_{eff}$=0.78 $\mu_B$ on Co sites of LaCo$_{0.78}$Sb$_2$, which is also consistent with the theoretical prediction qualitatively, but is a bit smaller value than the theoretical one (0.88$\mu_B$) in quantity. This is because the difference between the theoretical and the experimental findings originates from the presence of a deficiency of Co sites in the polycrystalline sample of LaCo$_{0.78}$Sb$_2$ in experiments, meaning that the hole-like charge carriers are doped into the system and reduce the value of the magnetic moment at the Co site (see details in Appendix~\ref{SecA4}), as well as the coherence length of long-range magnetic order. On the other hand, the magnetic moment at the Co site of LaCo$_{0.75}$Sb$_2$ is also evaluated using first-principles calculations. The calculated magnetic moment is 0.76 $\mu_B$, which is in good agreement with the experimental findings (see details in Appendix~\ref{SecA4}). Therefore, these experimental findings regarding the existence of AFM spin fluctuations further support the theoretical prediction of A-AFM ordering in LaCoSb$_2$.

\section{Symmetry analysis}\label{Sec4}

From the viewpoint of symmetry~\cite{Evarestov}, since the superconducting electrons in a superconductor are fermions, the total wave function needs to be antisymmetric under particle interchange, giving rise to the Pauli principles, which forbids identical fermions from sharing the same quantum states. In addition, the total wave function can be factored into space and spin components. For a Cooper pair with two electrons, the spin component of a pairing wave function can be characterized by its total spin $S=0$ (spin singlet states) or $S=1$ (spin triplet states). Obeying the total wave function to be antisymmetric, the spatial part of the wave function needs to have opposite symmetry. Therefore, the spin singlet state must have a symmetric spatial wave function. Similarly, spin triplet states have an antisymmetric spatial wave function.

For the classes of cuprates~\cite{Bednorz,ZXShen,CCTsuei} and iron-based superconductors~\cite{Kamihara,IIMazin,HDing,FWang}, the superconductivity emerges at the verge of AFM spin fluctuations, and the spatial wave function is symmetric, leading the pairing wave function to be a spin singlet. However, when the superconductivity emerges at the verge of FM spin fluctuations, the antisymmetric spatial wave function will result in spin triplet pairing for superconducting Cooper pairs.

Considering the existence of long-range FM order in the layered CoSb of LaCoSb$_2$, superconductivity might emerge at the verge of FM spin fluctuation when the charge carriers are gradually introduced into LaCoSb$_2$, making spin triplet pairings energetically favorable~\cite{Ogata}. Here it should be pointed out that superconductivity accompanied by an FM spin fluctuation has been observed experimentally in CoSb/SrTiO$_3$~\cite{CDing2019}, which supports the theoretical discussions. For spin triplet pairings, there are two possible candidates of $p$-wave and $f$-wave pairing gap structures up to the orbital angular momentum $L=3$, according to Ref.~\onlinecite{Sigrist}. In addition to the point symmetry of $D_{2d}$ for LaCoSb$_2$, the gap functions of superconducting doped LaCoSb$_2$ are described by the basis functions of different irreducible representations of the group $G=D_{2d}\times SO(3)$ in weak spin-orbit coupling, where $\times$ represents the direct product and $SO(3)$ represents all spin rotations~\cite{Annett}. Through the group analysis on the character table of $G$, gap structures of the chiral $p$ wave ($p_x+i p_y$) and chiral $f$ wave [$f_{x(x^2-3y^2)}+i f_{y(3x^2-y^2)}$] with time-reversal symmetry breaking are allowed for superconducting doped LaCoSb$_2$, where the possible non-unitary states are neglected, because these are usually energetically unfavorable in zero external magnetic field, noting that the $A_1$ phase of a non-unitary state in superfluid $^3$He is observed only in the presence of an applied magnetic field~\cite{Leggett,Wheatley}.

Furthermore, the symmetry of superconductivity is determined by the geometry of the Fermi surface~\cite{Ogata,JPHu}. From the calculated Fermi surface topologies shown in Fig.~\ref{fig1}, the fourfold symmetric structure leads to the chiral $p$-wave pairing being favorable energetically rather than the chiral $f$-wave pairing, which is usually favorable in threefold symmetric structures, such as the possible $f$-wave pairing proposed theoretically in the Na$_x$CoO$_2\cdot y$H$_2$O superconductor with a layered triangular lattice structure~\cite{Ogata,Johannes}. 

Thus, we propose the bulk material of LaCoSb$_2$ displaying strong layered FM order as a possible candidate parent compound of cobalt-based superconductors. When the FM order gets suppressed and superconductivity emerges as the charge carriers are introduced into the system, the possible chiral $p$-wave pairing state might be realized at the verge of FM spin fluctuation in doped superconducting LaCoSb$_2$.

\section{Conclusion}\label{Sec5}

Using first-principles calculations, we systemically study the electronic and magnetic structures of LaCoSb$_2$ in theory, and propose a weak AFM-ordered metal with FM order on CoSb layers as its ground state. This theoretical finding is further supported by our magnetization measurements on synthesized polycrystalline samples of LaCo$_{0.78}$Sb$_2$. These results suggest LaCoSb$_2$ as a possible candidate parent compound of cobalt-based superconductors, when the FM order gets suppressed and superconductivity emerges as the charge carriers are introduced into the systems. Since superconductivity might emerge at the verge of FM spin fluctuation, the symmetry requires doped LaCoSb$_2$ to a possible $p$-wave superconductor having potential applications in topological quantum computing in future.

\section*{ACKNOWLEDGMENTS}

This work was supported by the National Natural Science Foundation of China (Grant No. 11927807) and the Natural Science Foundation of Shanghai of China (Grant Nos. 18JC1420402 and 19ZR1402600). W. L. also acknowledges the start-up funding from Fudan University.

\begin{appendix}
\setcounter{figure}{0}
\renewcommand{\thefigure}{A\arabic{figure}}

\setcounter{table}{0}
\renewcommand{\thetable}{A\arabic{table}}

\section{The theoretical calculations of lattice stabilities}\label{SecA1}

\begin{figure}[tbp]
\centering{}\includegraphics[bb=15 25 490 380,width=8cm,height=6cm]{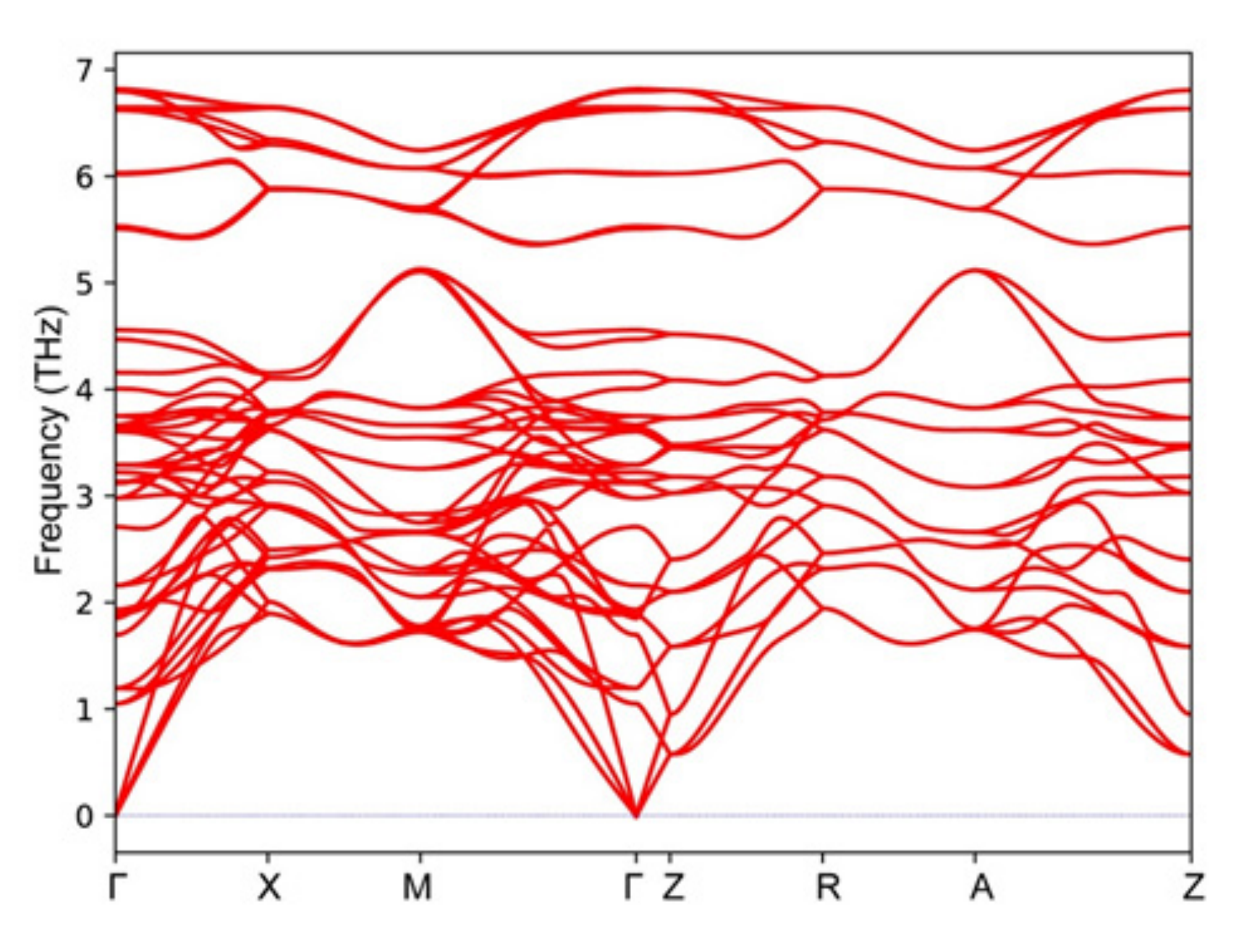}
\caption{(Color online) Phonon dispersion of LaCoSb$_2$ with the ground state of A-AFM order.}
\label{A1}
\end{figure}

To investigate the lattice dynamical stabilities of LaCoSb$_2$, we have performed the phonon properties calculations by using the finite difference method implemented in the Phonopy code~\cite{Togo2015}. A $3\times3\times1$ supercell with finite-difference displacements is used to calculate the force constants for the configuration of A-AFM order using the VASP code~\cite{Furthmuller} with an energy cutoff of $500$ eV for the plane-wave basis and an energy convergence criterion of $10^{-6}$ eV, which is the ground state of the parent compound LaCoSb$_2$. The phonon dispersion is shown in Fig.~\ref{A1}. It is clearly shown that the absence of imaginary frequencies in the phonon dispersion in Fig.~\ref{A1} demonstrates the stability of the crystal structure of LaCoSb$_2$.

\section{The orbital resolved energy bands of NM state of LaCoSb$_2$}\label{SecA2}

\begin{figure*}[tbp]
\centering{}\includegraphics[bb=15 25 2200 1200,width=18cm,height=9cm]{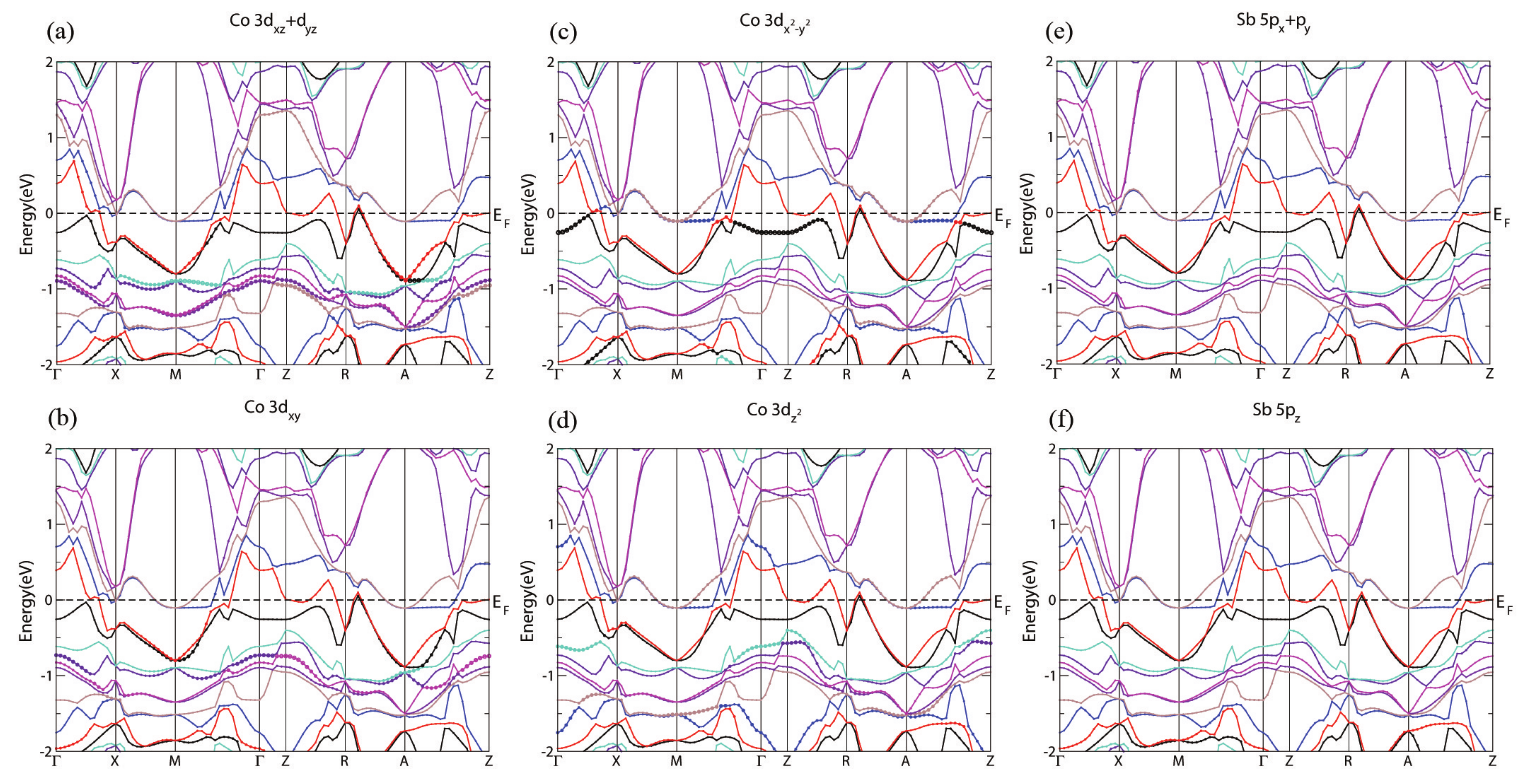}
\caption{(Color online) Projected Co $3d$ and Sb $5p$ orbital resolved energy bands of the NM state of LaCoSb$_2$. Fermi energies are set to zero.}
\label{A2}
\end{figure*}

We present the calculations on the orbital resolved energy bands of the NM state of LaCoSb$_2$. The projected Co $3d$ and Sb $5p$ orbitals resolved energy bands of the NM state of LaCoSb$_2$ are shown in Fig.~\ref{A2}.

\section{The orbital resolved energy bands of A-AFM state for LaCoSb$_2$}\label{SecA3}

\begin{figure*}[tbp]
\centering{}\includegraphics[bb=20 25 1455 2765,width=15cm,height=22cm]{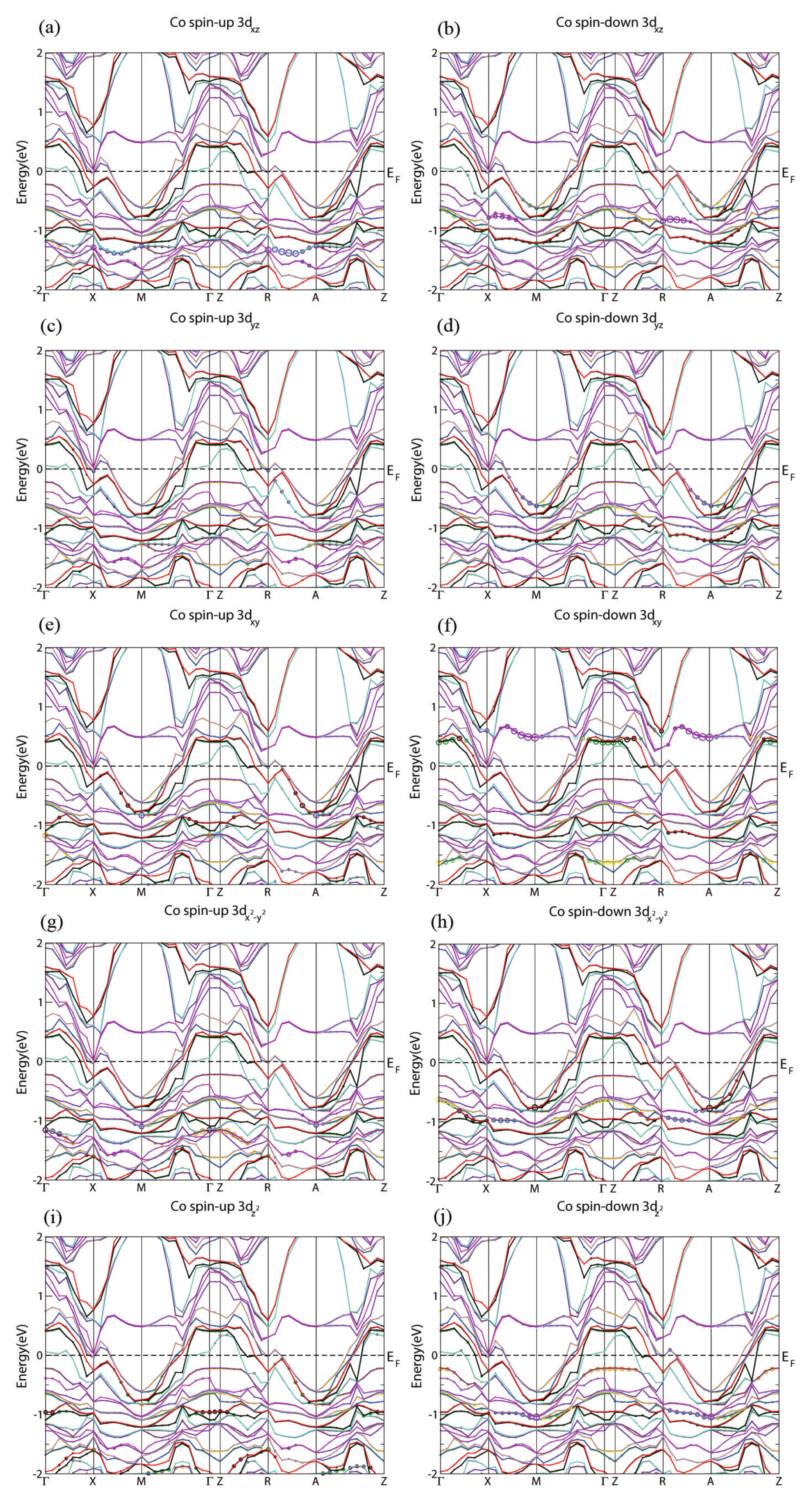}
\caption{(Color online) The $3d$ orbitals resolved energy bands projected on one of Co atoms in LaCoSb$_2$ with an A-AFM state. Fermi energies are set to zero.}
\label{A3}
\end{figure*}

For the A-AFM state, we also address the orbital resolved energy bands. The projected one of the Co $3d$ orbitals resolved energy bands of the A-AFM state are shown in Fig.~\ref{A3}. In comparison with the spin-up and spin-down species of Co $3d$ orbitals in the A-AFM state, the spin-dependent energy bands originating from the Co $t_{2g}$ orbitals change dramatically in involving the magnetic interaction, resulting in the opening of a gap among the $t_{2g}$ orbitals, and leaving the $e_g$ orbitals to electron conduction at the Fermi level in the A-AFM state.

\section{The XRD measurements on the polycrystalline samples of LaCo$_{1-x}$Sb$_2$}\label{SecA4}

\begin{figure}[tbp]
\centering{}\includegraphics[bb=15 25 560 480,width=8cm,height=6.5cm]{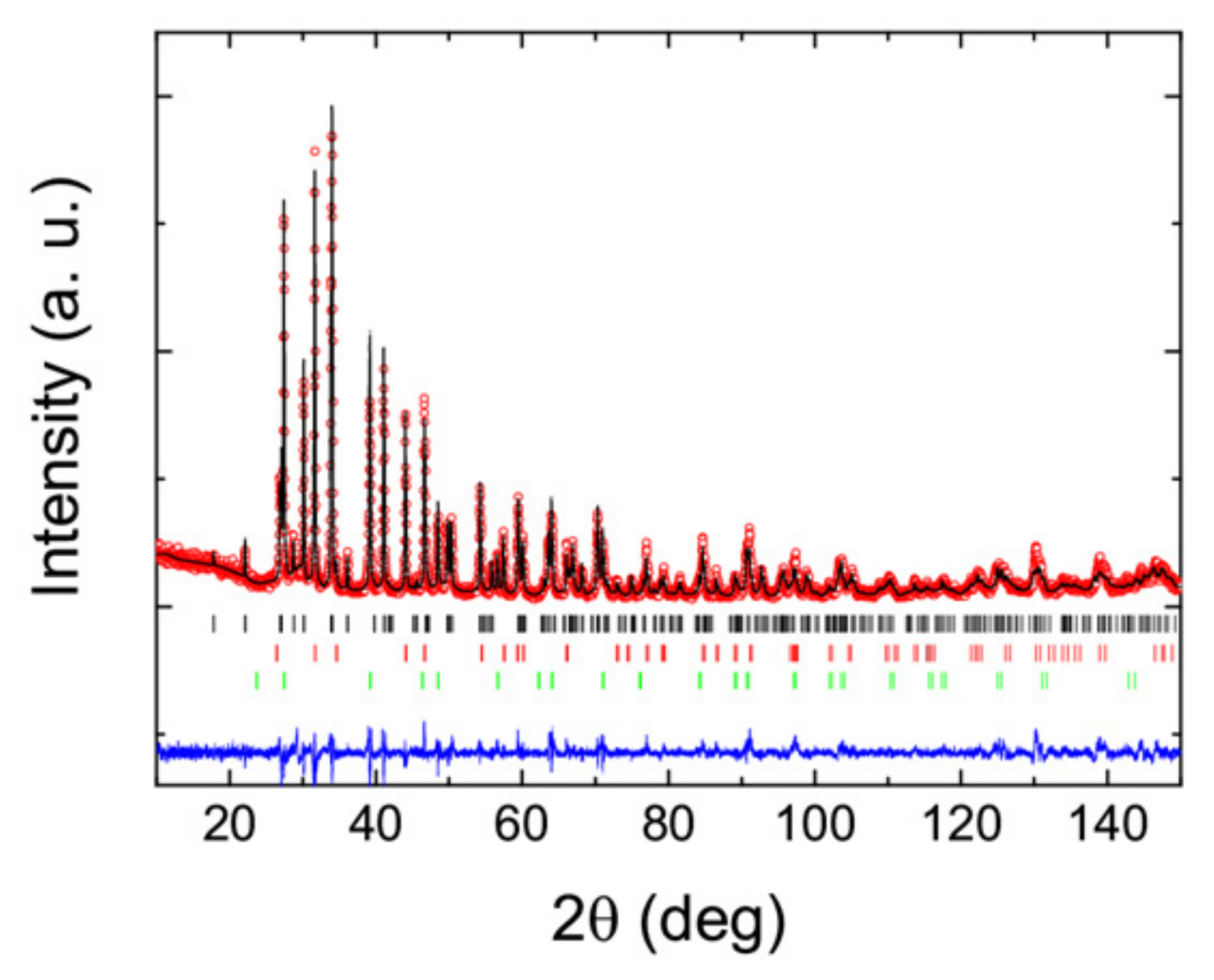}
\caption{(Color online) Observed (red circles) and calculated (solid black line) powder XRD patterns of LaCo$_{1-x}$Sb$_2$. The three rows of vertical bars show the calculated positions of Bragg reflections for LaCo$_{1-x}$Sb$_2$ (black), CoSb (red), and LaSb (green), respectively. The solid blue line at the bottom of the figure indicates the differences between observations and calculations.}
\label{A4}
\end{figure}

The crystal structure of the polycrystalline samples is systemically studied using a DX-2700 type powder x-ray diffractometer. The diffraction pattern is shown in Fig.~\ref{A4}. All the diffraction peaks can be grouped into the crystal structure of LaCo$_{1-x}$Sb$_2$ accompanied by some tiny impurity phases of CoSb and LaSb. It suggests a stable structure of the LaCoSb$_2$ phase at ambient pressure. The XRD data are also refined using the software of FullProf and the refinement pattern is shown in Fig.~\ref{A4} (see the solid black line). The refinement results for the main phase of LaCo$_{1-x}$Sb$_2$ are listed in Tables~\ref{tabelA1} and~\ref{tabelA2}. From these refinements, we find the deficiency of Co sites to be 22\%. This deficiency reduces the magnetic moment to 0.78 $\mu_B$ observed in the magnetization measurements, slightly smaller than the first-principles calculation of 0.88 $\mu_B$ for LaCoSb$_2$, since the deficiency of Co sites induces holelike charge carrier doping. On the other hand, the magnetic moment on the Co site of LaCo$_{0.75}$Sb$_2$ is also calculated using first-principles calculations. The calculated magnetic moment is 0.76 $\mu_B$, which is in good agreement with the experimental findings.

\begin{table}
\caption{Parameters for the structure refinement of LaCo$_{1-x}$Sb$_2$.}
\begin{ruledtabular} %
\begin{tabular}{ccccc}
Parameter  &  LaCo$_{1-x}$Sb$_2$   \\
\hline
Temperature    &  300 K \\
&\\
Wavelength     &  1.54\AA\\
&\\
R-factors      &  $R_p$=11.2\% \\
               &  $R_wp$=14.3\%\\
               &               \\
Crystal system &  Tetragonal   \\
               &\\
Space group    &P4/mmm (No. 129)\\
               &\\
Unit cell dimensions&  a=4.39174\AA, $\alpha$=90$^{\circ}$\\
&b=4.39174\AA, $\beta$=90$^{\circ}$\\
&c=9.94637\AA, $\gamma$=90$^{\circ}$\\
&\\
Volume&191.839\AA$^3$\\
\hline
Co vacancy    &  $x$=0.22\\
\end{tabular}\end{ruledtabular}
\label{tabelA1}
\end{table}

\begin{table}
\caption{Atomic coordinates of LaCo$_{1-x}$Sb$_2$.}
\begin{ruledtabular} %
\begin{tabular}{ccccc}
Atom  & $x$ & $y$ & $z$ & Occupancy   \\
\hline
La    & 0.25  & 0.25 &  0.25891  & 1 \\
\hline
Co    & 0.25  & 0.75 & 0.5       & 0.78 \\
\hline
Sb1   & 0.25  & 0.75 & 0.0       & 1 \\
\hline
Sb2   & 0.25  & 0.25 & 0.62314   & 1 \\
\end{tabular}\end{ruledtabular}
\label{tabelA2}
\end{table}

\end{appendix}


\begin{thebibliography}{10}

\bibitem{Qi2011} X.-L. Qi and S.-C. Zhang, {\it Topological insulators and superconductors}, Rev. Mod. Phys. \textbf{83}, 1057 (2011).

\bibitem{Flensberg} M. Leijnse and K. Flensberg, {\it Introduction to topological superconductivity and Majorana fermions}, Semicond. Sci. Technol. \textbf{27}, 124003 (2012).

\bibitem{Beenakker2013} C. W. J. Beenakker, {\it Search for Majorana Fermions in Superconductors}, Annu. Rev. Condens. Matter Phys. \textbf{4}, 113 (2013).

\bibitem{Sato2017} M. Sato and Y. Ando, {\it Topological superconductors: a review}, Rep. Prog. Phys. \textbf{80}, 076501 (2017).

\bibitem{Alicea} J. Alicea, {\it New directions in the pursuit of Majorana fermions in solid state systems}, Rep. Prog. Phys. \textbf{75}, 076501 (2012).

\bibitem{Franz2015} S. R. Elliott and M. Franz, {\it Majorana fermions in nuclear, particle, and solid-state physics}, Rev. Mod. Phys. 87, 137 (2015).

\bibitem{Kitaev} A. Y. Kitaev, {\it Fault-tolerant quantum computation by anyons}, Ann. Phys. (Amsterdam) \textbf{303}, 2 (2003).

\bibitem{Nayak2008} C. Nayak, S. H. Simon, A. Stern, M. Freedman, and S. Das Sarma, {\it Non-Abelian anyons and topological quantum computation}, Rev. Mod. Phys. \textbf{80}, 1083 (2008).

\bibitem{Ivanov2001} D. A. Ivanov, {\it Non-Abelian Statistics of Half-Quantum Vortices in $p$-Wave Superconductors}, Phys. Rev. Lett. \textbf{86}, 268 (2001).

\bibitem{Aguado} R. Aguado, {\it Majorana quasiparticles in condensed matter}, Riv. Nuovo Cimento \textbf{40}, 523 (2017).

\bibitem{Green2000} N. Read and D. Green, {\it Paired states of fermions in two dimensions with breaking of parity and time-reversal symmetries and the fractional quantum Hall effect}, Phys. Rev. B \textbf{61}, 10267 (2000).

\bibitem{Moore1991} G. Moore and N. Read, {\it Nonabelions in the fractional quantum Hall effect}, Nucl. Phys. B \textbf{360}, 362 (1991).

\bibitem{Read1992} N. Read and G. Moore, {\it Fractional quantum Hall effect and nonabelian statistics}, Prog. Theor. Phys. (Kyoto) Suppl. \textbf{107}, 157 (1992).

\bibitem{Maeno1994} Y. Maeno, H. Hashimoto, K. Yoshida, S. Nishizaki, T. Fujita, J. G. Bednorz, and F. Lichtenberg, {\it Superconductivity in a layered perovskite without copper}, Nature \textbf{372}, 532 (1994).

\bibitem{Maeno1998} K. Ishida, H. Mukuda, Y. Kitaoka, K. Asayama, Z. Q. Mao, Y. Mori, and Y. Maeno, {\it Spin-triplet superconductivity in Sr$_2$RuO$_4$ identified by $^{17}$O Knight shift}, Nature \textbf{396}, 658 (1998).

\bibitem{JXia2006} J. Xia, Y. Maeno, P. T. Beyersdorf, M. M. Fejer, and A. KapitulnikHigh, {\it Resolution polar Kerr effect measurements of Sr$_2$RuO$_4$: Evidence for broken time-reversal symmetry in the superconducting state}, Phys. Rev. Lett. \textbf{97}, 167002 (2006).

\bibitem{Kapitulnik2009} A. Kapitulnik, J. Xia, E. Schemm, and A. Palevski, {\it Polar Kerr effect as probe for time-reversal symmetry breaking in unconventional superconductors}, New Journal of Physics \textbf{11}, 055060 (2009).

\bibitem{Pustogow2019} A. Pustogow, Y. Luo, A. Chronister, Y.-S. Su, D. Sokolov, F. Jerzembeck, A. P. Mackenzie, C. W. Hicks, N. Kikugawa, S. Raghu, E. D. Bauer, and S. E. Brown, {\it Pronounced drop of $^{17}$O NMR Knight shift in superconducting state of Sr$_2$RuO$_4$}, arXiv: 1904.00047 (2019).

\bibitem{Ishida2019} K. Ishida, M. Manago, and Y. Maeno, {\it Reduction of the $^{17}$O Knight shift in the Superconducting State and the Heat-up Effect by NMR Pulses on Sr$_2$RuO$_4$}, arXiv: 1907.12236 (2019).

\bibitem{LFu2008} L. Fu and C. L. Kane, {\it Superconducting Proximity Effect and Majorana Fermions at the Surface of a Topological Insulator}, Phys. Rev. Lett. \textbf{100}, 096407 (2008).

\bibitem{Sau2010} J. D. Sau, R. M. Lutchyn, S. Tewari, and S. Das Sarma, {\it Generic New Platform for Topological Quantum Computation Using Semiconductor Heterostructures}, Phys. Rev. Lett. \textbf{104}, 040502 (2010).

\bibitem{Mourik2012} V. Mourik, K. Zuo, S. M. Frolov, S. R. Plissard, E. P. A. M. Bakkers, and L. P. Kouwenhoven, {\it Signatures of Majorana Fermions in Hybrid Superconductor-Semiconductor Nanowire Devices}, Science \textbf{336}, 1003 (2012).

\bibitem{MDeng2012} M. Deng, C. Yu, G. Huang, M. Larsson, P. Caroff, and H. Xu, {\it Anomalous zero-bias conductance peak in a Nb-InSb nanowire-Nb hybrid device}, Nano Lett. \textbf{12}, 6414 (2012).

\bibitem{ADas2012} A. Das, Y. Ronen, Y. Most, Y. Oreg, M. Heiblum, and H. Shtrikman, {\it Zero-bias peaks and splitting in an Al-InAs nanowire topological superconductor as a signature of Majorana fermions}, Nat. Phys. \textbf{8}, 887 (2012).

\bibitem{Deng2016} M. T. Deng, S. Vaitiek\`{e}nas, E. B. Hansen, J. Danon, M. Leijnse, K. Flensberg, J. Nyg{\aa}rd, P. Krogstrup, and C. M. Marcus, {\it Majorana bound state in a coupled quantum-dot hybrid-nanowire system}, Science \textbf{354}, 1557 (2016).

\bibitem{HZhang2018} H. Zhang, C.-X. Liu, S. Gazibegovic, D. Xu, J. A. Logan, G. Wang, N. van Loo, J. D. S. Bommer, M. W. A. de Moor, D. Car, R. L. M. Op het Veld, P. J. van Veldhoven, S. Koelling, M. A. Verheijen, M. Pendharkar, D. J. Pennachio, B. Shojaei, J. Sue Lee, C. J. Palmstr{\o}m, E. P. A. M. Bakkers, S. Das Sarma, and L. P. Kouwenhoven, {\it Quantized Majorana conductance}, Nature \textbf{556}, 74 (2018).

\bibitem{Perge2014} S. Nadj-Perge, I. K. Drozdov, J. Li, H. Chen, S. Jeon, J. Seo, A. H. MacDonald, B. A. Bernevig, and A. Yazdani, {\it Observation of Majorana fermions in ferromagnetic atomic chains on a superconductor}, Science \textbf{346}, 602 (2014).

\bibitem{JFJia2015} J.-P. Xu, M.-X. Wang, Z. L. Liu, J.-F. Ge, X. Yang, C. Liu, Z. A. Xu, D. Guan, C. L. Gao, D. Qian, Y. Liu, Q.-H. Wang, F.-C. Zhang, Q.-K. Xue, and J.-F. Jia, {\it Experimental detection of a Majorana mode in the core of a magnetic vortex inside a topological insulator-superconductor Bi$_2$Te$_3$/NbSe$_2$ heterostructure}, Phys. Rev. Lett. \textbf{114}, 017001 (2015).

\bibitem{JFJia2016} H.-H. Sun, K.-W. Zhang, L.-H. Hu, C. Li, G.-Y. Wang, H.-Y. Ma, Z.-A. Xu, C.-L. Gao, D.-D. Guan, Y.-Y. Li, C. Liu, D. Qian, Y. Zhou, L. Fu, S.-C. Li, F.-C. Zhang, and J.-F. Jia, {\it Majorana zero mode detected with spin selective Andreev reflection in the vortex of a topological superconductor}, Phys. Rev. Lett. \textbf{116}, 257003 (2016).

\bibitem{Banerjee2016} A. Banerjee, C. A. Bridges, J.-Q. Yan, A. A. Aczel, L. Li, M. B. Stone, G. E. Granroth, M. D. Lumsden, Y. Yiu, J. Knolle, S. Bhattacharjee, D. L. Kovrizhin, R. Moessner, D. A. Tennant, D. G. Mandrus, and S. E. Nagler, {\it Proximate Kitaev quantum spin liquid behaviour in a honeycomb magnet}, Nature Materials \textbf{15}, 733 (2016).


\bibitem{JXYin2015} J.-X. Yin, Z. Wu, J.-H. Wang, Z.-Y. Ye, J. Gong, X.-Y. Hou, L. Shan, A. Li, X.-J. Liang, X.-X. Wu, J. Li, C.-S. Ting, Z.-Q. Wang, J.-P. Hu, P.-H. Hor, H. Ding, and S. H. Pan, {\it Observation of a robust zero-energy bound state in iron-based superconductor Fe(Te,Se)}, Nat. Phys. \textbf{11}, 543 (2015).

\bibitem{DLFeng} Q. Liu, C. Chen, T. Zhang, R. Peng, Y.-J. Yan, C.-H.-P. Wen, X. Lou, Y.-L. Huang, J.-P. Tian, X.-L. Dong, G.-W. Wang, W.-C. Bao, Q.-H. Wang, Z.-P. Yin, Z.-X. Zhao, and D.-L. Feng, {\it Robust and Clean Majorana Zero Mode in the Vortex Core of High-Temperature Superconductor (Li$_{0.84}$Fe$_{0.16}$)OHFeSe}, Phys. Rev. X \textbf{8}, 041056 (2018).
\bibitem{HHWen} M. Chen, X. Chen, H. Yang, Z. Du, and H.-H. Wen, {\it Superconductivity with twofold symmetry in Bi$_2$Te$_3$/FeTe$_{0.55}$Se$_{0.45}$ heterostructures}, Science Advances \textbf{4}, eaat 1084 (2018).

\bibitem{Wang2018} D. Wang, L. Kong, P. Fan, H. Chen, S. Zhu, W. Liu, L. Cao, Y. Sun, S. Du, J. Schneeloch, R. Zhong, G. Gu, L. Fu, H. Ding, and H.-J. Gao, {\it Evidence for Majorana bound states in an iron-based superconductor}, Science \textbf{362}, 333 (2018).

\bibitem{SZhu2019} S. Zhu, L. Kong, L. Cao, H. Chen, S. Du, Y. Xing, W. Liu, D. Wang, C. Shen, F. Yang, J. Schneeloch, R. Zhong, G. Gu, L. Fu, Y.-Y. Zhang, H. Ding, and H.-J. Gao, {\it Nearly quantized conductance plateau of vortex zero mode in an iron-based superconductor}, Science \textbf{367}, 189 (2019).

\bibitem{CChen2020} C. Chen, K. Jiang, Y. Zhang, C. Liu, Y. Liu, Z. Wang, and J. Wang, {\it Atomic line defects and zero-energy end states in monolayer Fe(Te,Se) high-temperature superconductors}, Nature Physics online publication (2020).

\bibitem{Rogl1994} A. Leithe-Jasper and P. Rogl, {\it The crystal structure of NdFe$_{1-x}$Sb$_2$ and isotypic compounds RE(Fe, Co)$_{1-x}$Sb$_2$ (RE=La, Ce, Pr, Sm, Gd)}, Journal of Alloys and Compounds \textbf{203}, 133 (1994).

\bibitem{Katayama2013} N. Katayama, K. Kudo, S. Onari, T. Mizukami, K. Sugawara, Y. Sugiyama, Y. Kitahama, K. Iba, K. Fujimura, N. Nishimoto, M. Nohara, and H. Sawa, {\it Superconductivity in Ca$_{1-x}$La$_x$FeAs$_2$: A Novel 112-Type Iron Pnictide with Arsenic Zigzag Bonds}, J. Phys. Soc. Jpn. \textbf{82}, 123702 (2013).

\bibitem{Yakita2014} H. Yakita, H. Ogino, T. Okada, A. Yamamoto, K. Kishio, T. Tohei, Y. Ikuhara, Y. Gotoh, H. Fujihisa, K. Kataoka, H. Eisaki, and J.-i. Shimoyama, {\it A New Layered Iron Arsenide Superconductor: (Ca,Pr)FeAs$_2$}, J. Am. Chem. Soc. \textbf{136}, 846 (2014).

\bibitem{WLi} W. Li, S. Dong, C. Fang, and J. Hu, {\it Block antiferromagnetism and checkerboard charge ordering in the alkali-doped iron selenides $R_{1-x}$Fe$_{2-y}$Se$_2$}, Phys. Rev. B \textbf{85}, 100407(R) (2012).

\bibitem{WLi2012} W. Li, J.-X. Zhu, Y. Chen, and C. S. Ting, {\it First-principles calculations of the electronic structure of iron-pnictide EuFe$_2$(As,P)$_2$ superconductors: Evidence for antiferromagnetic spin order}, Phys. Rev. B \textbf{86}, 15519 (2012).

\bibitem{TZhang2019} T. Zhang, Y. Jiang, Z. Song, H. Huang, Y. He, Z. Fang, H. Weng, and C. Fang, {\it Catalogue of topological electronic materials}, Nature \textbf{566}, 475 (2019).

\bibitem{Zhang2020} W. Ding, J. Zeng, W. Qin, P. Cui, and Z. Zhang, {\it Exploring high transition temperature superconductivity in a freestanding or SrTiO$_3$-supported CoSb monolayer}, Phys. Rev. Lett. \textbf{124}, 027002 (2020).

\bibitem{CDing2019} C. Ding, G. Gong, Y. Liu, F. Zheng, Z. Zhang, H. Yang, Z. Li, Y. Xing, J. Ge, K. He, W. Li, P. Zhang, J. Wang, L. Wang, and Q.-K. Xue, {\it Signature of Superconductivity in Orthorhombic CoSb Monolayer Films on SrTiO$_3$(001)}, ACS Nano \textbf{13}, 10434 (2019).

\bibitem{QYWang2012} Q.-Y. Wang, Z. Li, W.-H. Zhang, Z.-C. Zhang, J.-S. Zhang, W. Li, H. Ding, Y.-B. Ou, P. Deng, K. Chang, J. Wen, C.-L. Song, K. He, J.-F. Jia, S.-H. Ji, Y.-Y. Wang, L.-L. Wang, X. Chen, X.-C. Ma, and Q.-K. Xue, {\it Interface-induced high-temperature superconductivity in single unit-cell FeSe films on SrTiO$_3$}, Chin. Phys. Lett. \textbf{29}, 037402 (2012).

\bibitem{Singh} D. J. Singh and L. Nordstrom, {\it Planewaves, Pseudopotentials, and the LAPW Method}, 2nd ed. (Springer-Verlag, Berlin, 2006), pp. 1C134.

\bibitem{wien2k} P. Blaha, K. Schwarz, G. Madsen, D. Kvasnicka, and J. Luitz, in WIEN2K, {\it An Augmented PlaneWave + Local Orbitals Program for Calculating Crystal Properties}, edited by K. Schwarz (Technical Univievsity Wien, Austria, 2001).

\bibitem{PBE1996} J. P. Perdew, K. Burke, and M. Erznerhof, {\it Generalized gradient approximation made simple}, Phys. Rev. Lett. \textbf{77}, 3865 (1996).

\bibitem{Li2014} W. Li, X.-Y. Wei, J.-X. Zhu, C. S. Ting, and Y. Chen, {\it Pressure-induced topological quantum phase transition in Sb$_2$Se$_3$}, Phys. Rev. B \textbf{89}, 035101 (2014).


\bibitem{Singh2008} D. J. Singh, {\it Electronic structure and doping in BaFe$_2$As$_2$ and LiFeAs: Density functional calculations}, Phys. Rev. B \textbf{78}, 094511 (2008).

\bibitem{WLi2012_EPL} W. Li, J. Li, J.-X. Zhu, Y. Chen, and C. S. Ting, {\it Pairing symmetry in iron-pnictide superconductor KFe$_2$As$_2$}, Europhysics Letters \textbf{99}, 57006 (2012).

\bibitem{Wollan} E. O. Wollan and W. C. Koehler, {\it Neutron diffraction study of the magnetic properties of the series of perovskite-type compounds [(1-x)La, xCa]MnO$_3$}, Phys. Rev. \textbf{100}, 545 (1955).

\bibitem{Dong2008} J. Dong, H. J. Zhang, G. Xu, Z. Li, G. Li, W. Z. Hu, D. Wu, G. F. Chen, X. Dai, J. L. Luo, Z. Fang, and N. L. Wang, {\it Competing orders and spin-density-wave instability in La(O$_{1-x}$F$_x$)FeAs}, Europhys. Lett. \textbf{83}, 27006 (2008).

\bibitem{Dai2008} C. de la Cruz, Q. Huang, J. W. Lynn, J. Li, W. Ratcliff II, J. L. Zarestky, H. A. Mook, G. F. Chen, J. L. Luo, N. L. Wang, and P. Dai, {\it Magnetic order close to superconductivity in the iron-based layered LaO$_{1-x}$F$_x$FeAs systems}, Nature \textbf{453}, 899 (2008).

\bibitem{ZZhou2019} Z. Zhou, W. T. Jin, W. Li, S. Nandi, B. Ouladdiaf, Z. Yan, X. Wei, X. Xu, W. H. Jiao, N. Qureshi, Y. Xiao, Y. Su, G. H. Cao, and Th. Br\"{u}ckel, {\it Universal critical behavior in the ferromagnetic superconductor Eu(Fe$_{0.75}$Ru$_{0.25}$)$_2$As$_2$}, Phys. Rev. B \textbf{100}, 060406(R) (2019).

\bibitem{Maekawa2004} S. Maekawa, T. Tohyama, S. E. Barnes, S. Ishihara, W. Koshibae, and G. Khaliullin, {\it Physics of Transition Metal Oxides}, (Springer-Verlag Berlin Heidelberg GmbH, 2004).

\bibitem{Mu2009} G. Mu, B. Zeng, X. Zhu, F. Han, P. Cheng, B. Shen, and H. H. Wen, {\it Synthesis, structural, and transport properties of the hole-doped superconductor Pr$_{1-x}$Sr$_{x}$FeAsO}, Phys. Rev. B \textbf{79}, 104501 (2009).

\bibitem{Amornpitoksuk2008} P. Amornpitoksuk, D. Ravot, A. Mauger, and J. C. Tedenac, {\it Structural and magnetic properties of the ternary solid solution between CoSb and Fe$_{1+\delta}$Sb}, Phys. Rev. B \textbf{77}, 144405 (2008).

\bibitem{bookQTM} N. Majlis, {\it The quantum theory of magnetism}, (World Scientific, 2007).

\bibitem{Evarestov} R. A. Evarestov and V. P. Smirnov, {\it Site symmetry in crystals}, (Springer-Verlag, Berlin, 1997).

\bibitem{Bednorz} J. G. Bednorz and K. A. M\"{u}ller, {\it Possible high T$_C$ superconductivity in the Ba-La-Cu-O system}, Zeitschrift f\"{u}r Physik B, \textbf{64}, 189 (1986).

\bibitem{ZXShen} Z.-X. Shen, D. S. Dessau, B. O. Wells, D. M. King, W. E. Spicer, A. J. Arko, D. Marshall, L. W. Lombardo, A. Kapitulnik, P. Dickinson, S. Doniach, J. DiCarlo, T. Loeser, and C. H. Park, {\it Anomalously large gap anisotropy in the a-b plane of Bi$_2$Sr$_2$CaCu$_2$O$_{8+\delta}$}, Phys. Rev. Lett. \textbf{70}, 1553 (1993).

\bibitem{CCTsuei} C. C. Tsuei, J. R. Kirtley, C. C. Chi, Lock See Yu-Jahnes, A. Gupta, T. Shaw, J. Z. Sun, and M. B. Ketchen, {\it Pairing symmetry and flux quantization in a tricrystal ring of superconducting YBa$_2$Cu$_3$O$_{7-\delta}$}, Phys. Rev. Lett. \textbf{73}, 593 (1994).

\bibitem{Kamihara} Y. Kamihara, T. Watanabe, M. Hirano, and H. Hosono, {\it Iron-based layered superconductor La[O$_{1-x}$F$_x$]FeAs ($x$ = 0.05-0.12) with T$_c$ = 26 K}, J. Am. Chem. Soc. \textbf{130}, 3296 (2008).

\bibitem{IIMazin} I. I. Mazin, D. J. Singh, M. D. Johannes, and M. H. Du, {\it Unconventional superconductivity with a sign reversal in the order parameter of LaFeAsO$_{1-x}$F$_x$}, Phys. Rev. Lett. \textbf{101}, 057003 (2008).

\bibitem{HDing} H. Ding, P. Richard, K. Nakayama, T. Sugawara, T. Arakane, Y. Sekiba, A. Takayama, S. Souma, T. Sato, T. Takahashi, Z. Wang, X. Dai, Z. Fang, G. F. Chen, J. L. Luo, and N. L. Wang, {\it Observation of Fermi-surface-dependent nodeless superconducting gaps in Ba$_{0.6}$K$_{0.4}$Fe$_2$As$_2$}, EuroPhysics Letters \textbf{83}, 47001 (2008).

\bibitem{FWang} F. Wang, H. Zhai, Y. Ran, A. Vishwanath, and D.-H. Lee, {\it Functional renormalization-group study of the pairing symmetry and pairing mechanism of the FeAs-based high-temperature superconductor}, Phys. Rev. Lett. \textbf{102}, 047005 (2009).

\bibitem{Ogata} Y. Tanaka, Y. Yanase, and M. Ogata, {\it Superconductivity in Na$_x$CoO$_2\cdot y$H$_2$O due to charge fluctuation}, Journal of the Physical Society of Japan \textbf{73}, 319 (2004).

\bibitem{Sigrist} M. Sigrist and K. Ueda, {\it Phenomenological theory of unconventional superconductivity}, Rev. Mod. Phys. \textbf{63}, 239 (1991).

\bibitem{Annett} J. F. Annett, {\it Symmetry of the order parameter for high-temperature superconductivity}, Advances in Physics \textbf{39}, 83 (1990).

\bibitem{Leggett} A. J. Leggett, {\it A theoretical description of the new phases of liquid $^3$He}, Rev. Mod. Phys. \textbf{47}, 331 (1975).

\bibitem{Wheatley} J. C. Wheatley, {\it Experimental properties of superfluid $^{3}$He}, Rev. Mod. Phys. \textbf{47}, 415 (1975).

\bibitem{JPHu} J. Hu and N. Hao, {\it $S_4$ symmetric microscopic model for iron-based superconductors}, Phys. Rev. X \textbf{2}, 021009 (2012).

\bibitem{Johannes} I. I. Mazin and M. D. Johannes, {\it Acritical assessment of the superconducting pairing symmetry in Na$_x$CoO$_2\cdot y$H$_2$O}, Nature Physics \textbf{1}, 91 (2005).

\bibitem{Togo2015} A. Togo and I. Tanaka, {\it First principles phonon calculations in materials science}, Scr. Mater. \textbf{108}, 1 (2015).

\bibitem{Furthmuller} G. Kresse and J. Furthmuller, {\it Efficient iterative schemes for ab initio total-energy calculations using a plane-wave basis set}, Phys. Rev. B \textbf{54}, 11169 (1996).



\end{thebibliography}
\end{document}